\documentstyle[12pt]{article}
 
\begin{document}

\begin{flushright} 
OCU-PHYS-168, 1997 
\end{flushright} 

\hspace*{1ex} 

\begin{center}
{\Large {\bf 
Comment on \lq\lq Infrared and pinching singularities in out of 
equilibrium QCD plasmas''}} 
\end{center}

\hspace*{3ex}

\hspace*{3ex}

\hspace*{3ex}

\begin{center} 
{\large {\sc A. Ni\'{e}gawa}\footnote{ 
E-mail: niegawa@sci.osaka-cu.ac.jp}

{\normalsize\em Department of Physics, Osaka City University } \\ 
{\normalsize\em Sumiyoshi-ku, Osaka 558, Japan} } \\
\end{center} 

\hspace*{2ex}

\hspace*{2ex}

\hspace*{2ex}

\hspace*{2ex}
\begin{center}
{\large {\bf Abstract}} \\ 
\end{center} 
\begin{quotation} 
Analyzing the dilepton production from out of equilibrium 
quark-gluon plasma, Le Bellac and Mabilat have recently pointed out 
that, in the reaction rate, the cancellation of mass (collinear) 
singularities takes place only in physical gauges, and not in 
covariant gauges. They then have estimated the contribution 
involving pinching singularities. After giving a general argument 
for the gauge independence of the production rate, we explicitly 
confirm the gauge independence of the mass-singular part. The 
contribution involving pinching singularities develops mass 
singularities, which is also gauge dependent. This \lq\lq 
additional'' contribution to the singular part is responsible for 
the gauge independence of the \lq\lq total'' singular part. We give 
a sufficient condition, under which cancellation of mass 
singularities takes place. 
\end{quotation} 
\newpage 
\setcounter{equation}{0} 
\def\theequation{\mbox{\arabic{equation}}} 
%% I %%%%%%%%%%%%%%%%%%%%%%%%%%%%
\noindent In the past years, much effort has been made to 
incorporate quantum field theory with nonequilibrium statistical 
mechanics, among which, we quote those of Altherr and Seibert 
\cite{alt-sei}, Altherr \cite{alt}, Baier et al \cite{bai}, Le 
Bellac and Mabilat \cite{lb} and the present author \cite{nie}. Out 
of equilibrium, pinching singularities appear \cite{alt-sei} in 
association with self-energy inserted propagator. It has been shown 
in \cite{alt} that resummation of self-energy part eliminates the 
pinching singularity (see also \cite{chou}). An application of this 
result to the rate of hard-photon production from nonequilibrium 
quark-gluon plasmas is made in \cite{bai}. A renormalization scheme 
of number density is introduced in \cite{nie}, such that the 
pinching singularities disappear. Le Bellac and Mabilat \cite{lb} 
are the first who have {\em explicitly} analyzed the infrared and 
mass (collinear) singularities in \lq\lq lepton-pair'' production 
rate. 

First of all, let us summarize the results of \cite{lb}. The 
production rate of a lepton pair from a quark-gluon plasma is 
proportional to $\Pi (Q) \equiv - i \Pi_{1 2} (Q)$, where $\Pi_{1 2} 
(Q)$ is the $(1, 2)$-component of the photon self-energy part in 
real-time massless QCD. Following \cite{lb}, we deal with $\Pi (Q)$ 
of a scalar \lq\lq photon.'' To two-loop order, $\Pi (Q)$ receives 
two contributions. The one $\Pi_\Sigma$ comes from the diagram with 
self-energy inserted quark propagator and the one $\Pi_V$ comes from 
the diagram with photon-quark vertex correction, 
%%%%% Equation %%%%%%%%%%%
\begin{eqnarray} 
\Pi (Q) & = & \Pi_V (Q) + \Pi_{\Sigma} (Q) \, , 
\label{wa} \\ 
\Pi_{\Sigma} (Q) & = & 2 i e^2 \int \frac{d^{\, 4} P}{(2 \pi)^4} 
\sum_{j, \, l = 1}^2 \mbox{Tr} \left[ S_{1 j} (P) \Sigma_{j l} (P) 
S_{l 2} (P) S_{2 1} (P - Q) \right] \, ,
\label{s1} \\ 
\Pi_V (Q) & = & - \frac{4}{3} e^2  g^2 \int 
\frac{d^{\, 4} P}{(2 \pi)^4} \int \frac{d^{\, 4} K}{(2 \pi)^4} 
g_{\mu \nu}^{(\mbox{\scriptsize{gauge}})} (K) 
\sum_{j, \, l = 1}^2 (-)^{j + l} \mbox{Tr} \left[ S_{1 j} 
(P - K) \gamma^\mu S_{j 2} (P) \right. \nonumber \\ 
& & \left. \times S_{2 l} (P - Q) \gamma^\nu S_{l 1} (P - Q - K) 
\right] \Delta_{l j} (K) \, , 
\label{v1} 
\end{eqnarray} 
%%% End %%%%%%%%%%%%%%%%%%
where 
%%%% Equation %%%%%%%%%%%%%%%%%%%%%%%%
\begin{eqnarray} 
\Sigma_{j l} (P) & = & i \frac{4}{3} g^2 (-)^{j + l} \int 
\frac{d^{\, 4} K}{(2 \pi)^4} 
g_{\mu \nu}^{(\mbox{\scriptsize{gauge}})} (K) 
\gamma^\mu S_{j l} (P - K) \gamma^\nu \Delta_{j l} (K) \, . 
\label{self} 
\end{eqnarray} 
%%% End %%%%%%%%%%%%%%%%%%
The (part of the) gluon propagator $\Delta_{l j} (K)$ takes the form 
\cite{lb} 
%%%% Equatioin %%%%%%%%%%%%%%%%%%%%%%%%%%%%%%%%%%%%%
\begin{eqnarray} 
\Delta_{1 1} (K) & = & \Delta^*_{2 2} (K) = (1 + f (K)) \Delta_R (K) 
+ f (K) \Delta_A (K) \, , \nonumber \\ 
\Delta_{1 2} (K) & = & f (K) (\Delta_R (K) + \Delta_A (K)) \, , 
\nonumber \\ 
\Delta_{2 1} (K) & = & ( 1 + f (K)) (\Delta_R (K) + \Delta_A (K)) 
\, , 
\label{hosi} 
\end{eqnarray} 
%%%%% End %%%%%%%%%%%%%%%%%%%%%%%%%
where $\Delta_{R, \, A} (K) = \pm i / (K^2 \pm i 0^+ \epsilon 
(k_0))$. The quark propagator $S_{l j} (K)$ takes the form $S_{l j} 
(K) = {K\kern-0.1em\raise0.3ex\llap{/}\kern0.15em\relax} 
\tilde{\Delta}_{l j} (K)$, where $\tilde{\Delta}_{l j} (K)$ is given 
by (\ref{hosi}) with the substitution $f \to - \tilde{f}$. $f$ 
($\tilde{f}$) is related to the distribution function of gluon $n$ 
(quark $\tilde{n}$) through 
%%% Equation %%%%%%%%%%%%%%%%%%%%%%%%%
\begin{eqnarray} 
f (K) & = & - \theta (- k_0) + \epsilon (k_0) \, n (|k_0|, \epsilon 
(k_0) \hat{{\bf k}}) \, , \nonumber \\ 
\tilde{f} (K) & = & \theta (- k_0) + \epsilon (k_0) \tilde{n} 
(|k_0|, \epsilon (k_0) \hat{{\bf k}}) \, , 
\label{bunpu} 
\end{eqnarray} 
%%% End %%%%%%%%%%%%%%%%%%%%%%%%%
where $\hat{{\bf k}} = {\bf k} / k$ with $k = |{\bf k}|$. Form of 
$g_{\mu \nu}^{(\mbox{\scriptsize{gauge}})} (K)$ in (\ref{v1}) and 
(\ref{self}) depends on the gauge choice: 
%%%% Equation %%%%%%%%%%%%%%%%%%%%%%%%%%%%%%%%%%%%%
\begin{eqnarray} 
g_{\mu \nu}^{(\mbox{\scriptsize{cov}})} (K) & = & g^{\mu \nu} - 
\eta \frac{K_\mu K_\nu}{K^2} \;\;\;\;\;\; \mbox{(covariant gauge)} 
\, , \label{cov} \\ 
g_{\mu \nu}^{(t)} (K) & = & g^{\mu \nu} - \frac{k_0}{k^2} (K_\mu 
n_\nu + n_\mu K_\nu) + \frac{K_\mu K_\nu}{k^2} \;\;\;\;\;\; 
\mbox{(Coulomb gauge)} \, , 
\label{t} 
\end{eqnarray} 
%%%% End %%%%%%%%%%%%%%%%%%%%%%%%%%
where $n^\mu = (1, {\bf 0})$. Observe that \cite{lb} 
%%%% Equation %%%%%%%%%%%%%%%%%
\begin{eqnarray} 
& & \sum_{j, \, l = 1}^2 S_{1 j} (P) \Sigma_{j l} (P) S_{l 2} (P) = 
F^{(n)} (P) + F^{(p)} (P) \, , 
\label{ide1} \\ 
& & \mbox{\hspace*{1.5ex}} F^{(n)} (P) = - \tilde{f} (P) ( 
\Delta_R^2 (P) - \Delta_A^2 (P)) 
{P\kern-0.1em\raise0.3ex\llap{/}\kern0.15em\relax} \, \mbox{Re} 
\Sigma_{11} (P) {P\kern-0.1em\raise0.3ex\llap{/}\kern0.15em\relax} 
\nonumber \\ 
& & \mbox{\hspace*{12.2ex}} - \frac{1}{2} \tilde{f} (P) ( \Delta_R^2 
(P) + \Delta_A^2 (P)) 
{P\kern-0.1em\raise0.3ex\llap{/}\kern0.15em\relax} 
(\Sigma_{1 2} (P) - \Sigma_{2 1} 
(P)) {P\kern-0.1em\raise0.3ex\llap{/}\kern0.15em\relax} \, , 
\label{f} \\ 
& & \mbox{\hspace*{1.5ex}} F^{(p)} (P) = 
\Delta_{\mbox{\scriptsize{R}}} (P) \Delta_{\mbox{\scriptsize{A}}} 
(P) {P\kern-0.1em\raise0.3ex\llap{/}\kern0.15em\relax} 
[ (1 - \tilde{f}(P) ) \Sigma_{1 2} 
(P) + \tilde{f} (P) \Sigma_{2 1} (P)] 
{P\kern-0.1em\raise0.3ex\llap{/}\kern0.15em\relax} \, . 
\label{p} 
\end{eqnarray} 
%%%% End %%%%%%%%%%%%%%%%%%%%%%%%
$F^{(n)}$ is the \lq\lq normal term,'' which is the counterpart of 
the one that is present in equilibrium thermal field theory (ETFT), 
while $F^{(p)}$ is the \lq\lq pinch term,'' which is absent in ETFT. 
Substituting (\ref{ide1}) - (\ref{p}) into (\ref{s1}), we have, with 
obvious notation, 
%%%% Equation %%%%%%%%%%%%%%%%%%%%%%%%%%%
\[ 
\Pi_{\Sigma} (Q) = \Pi_{\Sigma}^{(n)} (Q) + \Pi_{\Sigma}^{(p)} (Q) 
\, . 
\] 
%%%% End %%%%%%%%%%%%%
Le Bellac and Mabilat \cite{lb} have shown that, in Coulomb gauge, 
the mass singularities cancel out both in $\Pi_{\Sigma}^{(n)} (Q)$ 
and $\Pi_V (Q)$, Eq. (\ref{v1}). While, in the covariant gauge, the 
cancellation holds only for $\Pi_{\Sigma}^{(n)} (Q)$, and in 
$\Pi_V (Q)$ there survives mass singularity. Then, the authors of 
\cite{lb} have concluded that whether or not the singularity 
cancellation takes place is gauge dependent. [In Appendix A, we show 
in a gauge-independent manner how the cancellations of mass 
singularities take place in $\Pi_\Sigma^{(n)} (Q)$, re\-confirming 
the result of \cite{lb}.] 

We first verify that $\Pi (Q)$, Eq. (\ref{wa}), is gauge 
independent. Proof goes just as in vacuum ($T = 0$) theory. Consider 
the difference 
%%%% Equation %%%%%%%%%%%%%%%%%%%%%%%%%%%%%%%%%%%%%%%%
\begin{eqnarray} 
\delta \Pi (Q) & \equiv & \Pi_\Sigma (Q) \mbox{\hspace*{.4mm}} 
\rule[-2.8mm]{.14mm}{6.2mm} \raisebox{-2.85mm}{\scriptsize{$\; 
\mbox{\scriptsize{covariant}}$}} - \Pi_\Sigma (Q) 
\mbox{\hspace*{.4mm}} \rule[-2.8mm]{.14mm}{6.2mm} 
\raisebox{-2.85mm}{\scriptsize{$\; \mbox{\scriptsize{Cou}}$}} \, , 
\end{eqnarray} 
%%%% End %%%%%%%%%%%%%%%%%%%%%%%%%
where \lq\lq Cou'' stands for Coulomb. Observe that, from 
(\ref{cov}) and (\ref{t}), the difference 
$g_{\mu \nu}^{(\mbox{\scriptsize{cov}})} (K) - g_{\mu \nu}^{(t)} 
(K)$ is proportional to $K_\mu$ and/or $K_\nu$. Then, in 
evaluating $\delta \Pi (Q)$, we can use Ward-Takahashi relation, 
%%%% Equation %%%%%%%%%%%%%%%%%
\[ 
S_{j k} (P - K) {K\kern-0.1em\raise0.3ex\llap{/}\kern0.15em\relax} 
S_{k l} (P) = - i (-)^k \delta_{k l} S_{j k} (P - K) 
+ i (-)^j \delta_{j k} S_{k l} (P) \, , 
\] 
%%%% End %%%%%%%%%%%%%%%%%%%%%%%%
with no summation over repeated indices. After doing this, we see 
that, among many terms in the resultant expression for $\delta \Pi 
(Q)$, complete cancellations occur, so that $\delta \Pi (Q)$ 
vanishes and then is, of course, free from mass singularities. 

On the light of the above observation, let us make a closer 
inspection of the results of \cite{lb}. As a covariant gauge, as in 
\cite{lb}, we take the Feynman gauge ($\eta = 0$ in (\ref{cov})) 
throughout in the sequel. We analyze $\Pi_\Sigma (Q)$, Eq. 
(\ref{s1}), which includes ${\cal G}^{\mu \nu} \equiv 
{P\kern-0.1em\raise0.3ex\llap{/}\kern0.15em\relax} \gamma^\mu 
( {P\kern-0.1em\raise0.3ex\llap{/}\kern0.15em\relax} - 
{K\kern-0.1em\raise0.3ex\llap{/}\kern0.15em\relax} ) \gamma^\nu 
{P\kern-0.1em\raise0.3ex\llap{/}\kern0.15em\relax}$. Simple algebra 
yields 
%%%%% Equation %%%%%%%%%%%%%%%%
\begin{eqnarray} 
g_{\mu \nu} \, {\cal G}^{\mu \nu} & = & - 2 P^2 
{K\kern-0.1em\raise0.3ex\llap{/}\kern0.15em\relax} \, , 
\label{pre} \\ 
\delta g_{\mu \nu} \, {\cal G}^{\mu \nu} & = & 2 P^2 \frac{k_0}{k^2} 
\left[ (2 p_0 - k_0) 
{P\kern-0.1em\raise0.3ex\llap{/}\kern0.15em\relax} - p_0 
{K\kern-0.1em\raise0.3ex\llap{/}\kern0.15em\relax} \right] 
+ O ((P^2)^2) \, , 
\label{tr} 
\end{eqnarray} 
%%%% End %%%%%%%%%%%%%%%%%%%%%%%%
where $\delta g_{\mu \nu} \equiv g_{\mu \nu} - g_{\mu \nu}^{(t)} 
(K)$ and use has been made of $K^2 = (P - K)^2 = 0$. Thus, both 
$g_{\mu \nu} {\cal G}^{\mu \nu}$ and $\delta g_{\mu \nu} {\cal 
G}^{\mu \nu}$ are proportional to $P^2$. Now we notice that $P^2 
\Delta_{\mbox{\scriptsize{R}}} (P) \Delta_{\mbox{\scriptsize{A}}} 
(P)$ $= {\bf P} / P^2$. This means that the pinching singularity in 
the \lq\lq pinch'' term $F^{(p)}$, Eq. (\ref{p}), turns out to be a 
mass singularity. The $(P^2)^2$ term in (\ref{tr}) does not lead to 
mass-singular contribution. As in \cite{lb}, let us restrict our 
concern to singular contributions and ignore the $(P^2)^2$ term. 

As far as mass-singular contributions are concerned, above 
observation $\delta \Pi (Q) = 0$ tells us that $\Pi_V^{\, 
\mbox{\scriptsize{sing}}} \mbox{\hspace*{.4mm}} 
\rule[-2.0mm]{.14mm}{5.7mm} \raisebox{-1.95mm}{\scriptsize{$\; 
\mbox{\scriptsize{Fey}}$}}$ $= - \delta \Pi_\Sigma^{(p), \, 
\mbox{\scriptsize{sing}}}$, where \lq\lq Fey'' stands for Feynman. 
$\Pi_V^{\, \mbox{\scriptsize{sing}}} \mbox{\hspace*{.4mm}} 
\rule[-2.0mm]{.14mm}{5.7mm} \raisebox{-1.95mm}{\scriptsize{$\; 
\mbox{\scriptsize{Fey}}$}}$ with $Q = (q_0, {\bf q} = {\bf 0})$ is 
explicitly evaluated in \cite{lb}. As a check, in Appendix B, we 
evaluate $\delta \Pi_\Sigma^{(p), \, \mbox{\scriptsize{sing}}} 
(q_0, {\bf 0})$ and confirm\footnote{There is a missing term in 
$\Pi_V \mbox{\hspace*{.5mm}} \rule[-2.0mm]{.14mm}{5.0mm} 
\raisebox{-1.95mm}{\scriptsize{$\; \mbox{\scriptsize{Fey}}$}}$ 
in \cite{lb} (see Appendix B).} $\delta \Pi_\Sigma^{(p), \, 
\mbox{\scriptsize{sing}}} = - \Pi_V^{\, \mbox{\scriptsize{sing}}} 
\mbox{\hspace*{.4mm}} \rule[-2.0mm]{.14mm}{5.7mm} 
\raisebox{-1.95mm}{\scriptsize{$\; \mbox{\scriptsize{Fey}}$}}$. 

The singular part of $\Pi (q_0 \equiv 2 \kappa, {\bf 0})$, being 
gauge independent, is also evaluated in Appendix B, 
%%%%  Equation %%%%%%%%%%%%%%%%%
\begin{eqnarray} 
\Pi^{\, \mbox{\scriptsize{sing}}} & = & \Pi_\Sigma^{(p), \, 
\mbox{\scriptsize{sing}}} \mbox{\hspace*{.4mm}} 
\rule[-3mm]{.14mm}{8.5mm} \raisebox{-2.85mm}{\scriptsize{$\; 
\mbox{\scriptsize{Cou}}$}} = \Pi_V^{\, \mbox{\scriptsize{sing}}} 
\mbox{\hspace*{.4mm}} \rule[-3mm]{.14mm}{8.5mm} 
\raisebox{-2.85mm}{\scriptsize{$\; \mbox{\scriptsize{Fey}}$}} + 
\Pi_\Sigma^{(p), \, \mbox{\scriptsize{sing}}} \mbox{\hspace*{.4mm}} 
\rule[-3mm]{.14mm}{8.5mm} \raisebox{-2.85mm}{\scriptsize{$\; 
\mbox{\scriptsize{Fey}}$}} \nonumber \\ 
& & = 
- \frac{32}{3 \pi} \alpha \alpha_s \kappa^2 \ln 
\frac{1}{\epsilon_y} \int \frac{d \Omega_{\hat{\bf k}}}{4 \pi} \tilde{n} 
(\kappa, \hat{{\bf k}}) \tilde{n} (\kappa, - \hat{{\bf k}}) \left[ 
\int_{\epsilon_z}^1 d z \, \frac{(z - 1)^2 + 1}{2 z} \right. 
\nonumber \\ 
& & \left. 
+ \int_0^\infty dz \frac{z^2 + 2}{z} n (\kappa z, \hat{{\bf k}}) 
+ \int_0^\infty d z \, {\bf P} \frac{z (z^2 + 1)}{z^2 - 
1} \tilde{n} (\kappa z, \hat{{\bf k}}) \right] \nonumber \\ 
& & + \frac{16}{3 \pi} \alpha \alpha_s \kappa^2 \ln 
\frac{1}{\epsilon_y} \int \frac{d \Omega_{\hat{\bf k}}}{4 \pi} \tilde{n} 
(\kappa, - \hat{{\bf k}}) \left[ \int_{\epsilon_z}^1 dz 
\frac{(z - 1)^2 + 1}{z} 
n (\kappa z, \hat{{\bf k}}) \tilde{n} (\kappa (1 - z), 
\hat{{\bf k}}) \right. \nonumber \\ 
& & + \int_1^\infty dz \frac{(z - 1)^2 + 1}{z} 
n (\kappa z, \hat{{\bf k}}) (1 - \tilde{n} (\kappa 
(z - 1), \hat{{\bf k}}) \nonumber \\ 
& & \left. + \int_{\epsilon_z}^\infty dz 
\frac{(z + 1)^2 + 1}{z} 
(1 + n (\kappa z, \hat{{\bf k}}) ) \tilde{n} 
(\kappa (1 + z), \hat{{\bf k}}) \right] \, . 
\label{final} 
\end{eqnarray} 
%%%% End %%%%%%%%%%%%%%%%%%%%%%%%
Here $d \Omega_{\hat{{\bf k}}}$ is the element of the solid angle in 
the ${\bf k}$-space. The cutoff factor $\epsilon_y$ is defined by 
$1 - |\hat{{\bf p}} \cdot \hat{{\bf k}}| \geq \epsilon_y$ (cf. Eq. 
(\ref{A3}) in Appendix A) and $\epsilon_z$ is the infrared cutoff $k 
\geq \epsilon_z \kappa$. 

Let us clarify the relation between the present result and the 
result of \cite{lb}. We start with picking out from $F^{(p)} (P)$, 
%%%%%  Equation %%%%%%%%%%%%%%%%%%%%%%%%%%%%
\begin{equation} 
\overline{\Sigma} (P) \equiv (1 - \tilde{f}(P)) \Sigma_{1 2} (P) + 
\tilde{f} (P) \Sigma_{2 1} (P) \, . 
\label{sankaku} 
\end{equation} 
%%%% End %%%%%%%%%%%%%%%%%%%%%%%%%%%
For the purpose of estimating $\Pi_\Sigma^{(p)}$, Le Bellac and 
Mabilat \cite{lb} have analyzed $\overline{\Sigma} (P)$ within 
the hard-thermal-loop resummation scheme. The net production rate of 
an (anti)quark is given by $\overline{\Gamma} (P) = \mbox{Tr} [ i 
\overline{\Sigma} (P) 
{P\kern-0.1em\raise0.3ex\llap{/}\kern0.15em\relax} ] / (4 p)$, with 
$P = (p, {\bf p})$ for quark and $P = (- p, - {\bf p})$ for 
antiquark. Arguing that $\overline{\Gamma} (P)$ on the mass shell 
$P^2 = 0$, being gauge independent, is relevant to $\Pi_\Sigma 
(Q)$, the authors of \cite{lb} have concluded that $\Pi (Q)$ is 
gauge dependent since $\Pi^{\, \mbox{\scriptsize{sing}}} (Q)$ is. 
It is clear from the above argument that this is not the case. As 
has been discussed above in conjunction with (\ref{tr}), 
${P\kern-0.1em\raise0.3ex\llap{/}\kern0.15em\relax} \delta 
\overline{\Sigma} (P) 
{P\kern-0.1em\raise0.3ex\llap{/}\kern0.15em\relax}$ (as well as 
${P\kern-0.1em\raise0.3ex\llap{/}\kern0.15em\relax} 
\overline{\Sigma} (P) 
{P\kern-0.1em\raise0.3ex\llap{/}\kern0.15em\relax} 
\mbox{\hspace*{.4mm}} 
\rule[-2.0mm]{.14mm}{5.7mm} \raisebox{-1.95mm}{\scriptsize{$\; 
\mbox{\scriptsize{Fey}}$}}$) vanishes on the mass shell $P^2 = 
0$. [As a matter of fact $\Sigma_{1 2 (2 1)} (P)$ in (\ref{sankaku}) 
vanishes on the mass shell $P^2 = 0$, since $P^2$, $K^2$ and 
$(P - K)^2$ cannot vanish simultaneously.] However, as noted above, 
in calculating $\delta F^{(p)}$ (cf. Eq. (\ref{p})), the factor 
$P^2$, Eq. (\ref{tr}), \lq\lq eliminates'' one $\Delta$ and the 
mass-singular contribution $\delta \Pi_\Sigma^{(p), \, 
\mbox{\scriptsize{sing}}}$ (as well as $\Pi_\Sigma^{(p), \, 
\mbox{\scriptsize{sing}}} \mbox{\hspace*{.4mm}} 
\rule[-2.0mm]{.14mm}{5.7mm} \raisebox{-1.95mm}{\scriptsize{$\; 
\mbox{\scriptsize{Fey}}$}}$) emerges. Thus we have learned that the 
mass-singular contribution does not come from the gauge-independent 
quantity, 
%%%%%%%%% Equation %%%%%%%%%%
\[ 
{P\kern-0.1em\raise0.3ex\llap{/}\kern0.15em\relax} 
\overline{\Sigma} (P) 
{P\kern-0.1em\raise0.3ex\llap{/}\kern0.15em\relax} 
\mbox{\hspace*{.4mm}} \rule[-2.5mm]{.14mm}{5.9mm} 
\raisebox{-2.55mm}{\scriptsize{$\; P^2 = 0$}} = - 2 i p 
\overline{\Gamma} (P) 
{P\kern-0.1em\raise0.3ex\llap{/}\kern0.15em\relax} 
\mbox{\hspace*{.4mm}} \rule[-2.5mm]{.14mm}{5.9mm} 
\raisebox{-2.55mm}{\scriptsize{$\; P^2 = 0$}} \; (= 0) \, , 
\] 
%%%%% End %%%%%%%%%%
but comes from the gauge-dependent quantity 
%%%%%%%%% Equation %%%%%%%%%%
\[ 
d 
{P\kern-0.1em\raise0.3ex\llap{/}\kern0.15em\relax} \overline{\Sigma} 
(P) {P\kern-0.1em\raise0.3ex\llap{/}\kern0.15em\relax} / d P^2 
\mbox{\hspace*{.4mm}} \rule[-2.5mm]{.14mm}{5.9mm} 
\raisebox{-2.55mm}{\scriptsize{$\; P^2 = 0$}} = 
{P\kern-0.1em\raise0.3ex\llap{/}\kern0.15em\relax} \overline{\Sigma} 
(P) {P\kern-0.1em\raise0.3ex\llap{/}\kern0.15em\relax} / P^2 
\mbox{\hspace*{.4mm}} \rule[-2.5mm]{.14mm}{5.9mm} 
\raisebox{-2.55mm}{\scriptsize{$\; P^2 = 0$}} \, . 
\]  
%%%%%%%% End %%%%%%%%%%%%%%%%%%%%%

The observation made above in conjunction with (\ref{tr}) applies to 
the contribution (to $\delta F^{(p)}$) from the soft-$K$ region (cf. 
(\ref{p}) with (\ref{self})), in which $K^2 \neq 0$. [The soft 
$(P - K)$-region is not important, at least, for the system, which 
is not far from thermal and chemical equilibrium.] Above observation 
on $\delta \Pi_\Sigma^{(p), \, \mbox{\scriptsize{sing}}}$ holds as 
it is, except that $\Sigma_{1 2 (2 1)} (P)$ does not vanish on the 
mass shell. It is to be noted in passing that $\Pi_\Sigma^{(p)} 
\mbox{\hspace*{.4mm}} \rule[-2.0mm]{.14mm}{5.7mm} 
\raisebox{-1.95mm}{\scriptsize{$\; \mbox{\scriptsize{Fey}}$}}$ 
develops pinch singularity. This is because, in the present case, 
$K^2 \neq 0$, we have, in place of (\ref{pre}), $g_{\mu \nu} 
{\cal G}^{\mu \nu} = - 2 P^2 
{K\kern-0.1em\raise0.3ex\llap{/}\kern0.15em\relax} + 2 K^2 
{P\kern-0.1em\raise0.3ex\llap{/}\kern0.15em\relax}$. The second term 
on the right-hand side leads to pinching singularity in 
$\Pi_\Sigma^{(p)}$. Because of the factor $K^2$, which is small, the 
\lq\lq residue'' of the pinching contribution is relatively small. 

Finally, for an instructive purpose, we derive a sufficient 
condition for the absence of mass singularities, i.e., the condition 
under which (\ref{final}) vanishes. The condition is 
%%%%  Equation %%%%%%%%%%%%%%%%%
\begin{equation} 
[1 + f (K)] [1 - \tilde{f} (P - K)] = 
[1 - \tilde{f} (P)] [1 + f (K) - \tilde{f} (P - K)] \, , 
\label{cond1} 
\end{equation} 
%%%%% End %%%%%%%%%%%%%%
which holds for equilibrium case. For simplicity of presentation, we 
take the self-interacting scalar theory counterpart \cite{lb} to 
(\ref{cond1}), 
%%%%  Equation %%%%%%%%%%%%%%%%%
\begin{equation} 
[1 + f (K)] [1 + f (P - K)] = 
[1 + f (P)] [1 + f (K) + f (P - K)] \, . 
\label{cond2} 
\end{equation} 
%%%% End %%%%%%%%%%%%%%%
The same conclusion, obtained below, applies to (\ref{cond1}). 

We start with the following observation. It is well known that the 
reaction rates are free from mass singularities if all the (energy) 
degenerate states are prepared in the initial state (see e.g., 
\cite{nie1})). Mass singularities arise from collinear 
configurations of massless particles. For convenience, we enclose 
the system into a box of volume $V \; (= L^3)$ and employ a periodic 
boundary condition. In the interaction representation, the scalar 
field may be expressed as a superposition of plane waves, whose 
\lq\lq coefficients'' $a_{\hat{\bf k}_j}$ and 
$a^\dagger_{\hat{\bf k}_j}$ are the annihilation and creation 
operators, respectively. Here $\hat{{\bf k}_j}$ is the momenta of a 
particle with mode $j$. The, above observation tells us that for the 
system characterized by the density operator 
%%%%%  Equation %%%%%%%%%%%%%%%%%%%%%%%%%%%%
\begin{equation} 
\rho \left( \sum_j g (\hat{{\bf k}}_j) k_j a^\dagger_{\hat{\bf k}_j} 
a_{\hat{\bf k}_j} \right) 
\label{density} 
\end{equation} 
%%%% End %%%%%%%%%%%
reaction rates are free from mass singularities. In (\ref{density}), 
$k_j = | {\bf k}_j |$ is the energy of a particle with ${\bf k}_j$ 
and $f ({\hat{\bf k}}_j)$ is a (smooth) function. If $f 
({\hat{\bf k}}_j) = 1$, the argument of $\rho$ in (\ref{density}) is 
essentially an energy operator in the initial (remote past) state. 
It should be remarked that, even in the present context, the form of 
the density operator takes much more complicated form, since the 
system is in general not uniform in space-time. However, as far as a 
reaction rate taking place in the system is concerned, all the 
propagators describing the reaction rate have \cite{lb,chou} common 
center-of-mass or macroscopic coordinates of the space-time region, 
where the reaction takes place. Then it is sufficient to use 
(\ref{density}). 

Let us show that (\ref{density}) leads to (\ref{cond2}). We start 
with computing $\mbox{Tr} \, \rho$ 
%%%%%  Equation %%%%%%%%%%%%%%%%%%%%%%%%%%%%
\begin{eqnarray*} 
\mbox{Tr} \, \rho & = & \sum_{\{ n_j \}} \rho \left( \sum_j k_j \, g 
(\hat{{\bf k}}_j) n_{\hat{\bf k}_j} \right) \nonumber \\ 
& = & \int d E \, \rho (E) \sum_{\{ n_j \}} \delta \left( E - 
\sum_j k_j \, g (\hat{{\bf k}}_j) n_{\hat{\bf k}_j} \right) 
\nonumber \\ 
& = & \int d E \, \rho (E) \sum_{\{ n_j \}} \int_{- \infty}^\infty 
\frac{d \xi}{2 \pi} e^{i \xi E} e^{- i \xi \sum_j k_j \, g 
(\hat{{\bf k}}_j) n_{\hat{\bf k}_j}} \nonumber \\ 
& = & \int d E \, \rho (E) \int_{- \infty}^\infty 
\frac{d \xi}{2 \pi} e^{i \xi E} {\cal G} \, , 
\end{eqnarray*} 
%%%%% End %%%%%%%%%%
where 
%%%%%% Equation %%%%%%%%%%%%%%%%%%%%%%%%%%%
\[ 
{\cal G} \equiv \prod_j \frac{1}{1 - e^{- i \xi k_j g 
(\hat{{\bf k}}_j) }} \, . 
\] 
%%%% End %%%%%%%%%%%
Now, the number density is evaluated as 
%%%% Equation %%%%%%%%%%%%%%%%%%%%%%%%%%%%%
\begin{eqnarray*} 
n (k_l, \hat{{\bf k}}_l) & \equiv & \mbox{Tr} \, n_{{\hat{\bf k}}_l} 
\rho \nonumber \\ 
& = & \int d E \, \rho (E) \int \frac{d \xi}{2 \pi} e^{i \xi E} 
\frac{1}{e^{i \xi k_l \, g ({\hat{\bf k}}_l) } - 1} \, {\cal G} \, . 
\end{eqnarray*} 
%%%% End %%%%%%%%%%%%%%%%%%%%%%%%%%
Armed with the above preliminary, we take (\ref{cond2}) with $p_0 > 
k_0 > 0$. The right-hand side of (\ref{cond2}) reads\footnote{For 
the mode-overlapping \lq\lq points'', ${\bf k}_n = {\bf k}_j$, 
${\bf k}_n = {\bf 0}$, ${\bf k}_j = {\bf 0}$, (\ref{en}) does not 
hold in general. The contributions from such \lq\lq points'' are $O 
(1 / V)$ smaller than the \lq\lq reference contribution'' and vanish 
in the limit $V \to \infty$. Incidentally, for the case of 
equilibrium system, (\ref{en}) holds \cite{nie2}.} 
%%%%% Equation %%%%%%%%%%%%%%%%%%%%%%%%%%%%
\begin{eqnarray} 
& & [1 + n (k_n, \hat{{\bf k}}_n)] [1 + n (k_j, \hat{{\bf k}}_j) + n 
(|{\bf k}_n - {\bf k}_j|, \widehat{{\bf k}_n - {\bf k}_j} ) ] 
\nonumber \\ 
& & \mbox{\hspace*{4ex}} = \int d E \, \rho (E) 
\int_{- \infty}^\infty \frac{d \xi}{2 \pi} e^{i \xi E} [ 1 + 
N (\xi, {\bf k}_n) ] [ 1 +  N (\xi, {\bf k}_j) + N (\xi, 
{\bf k}_n - {\bf k}_j ) ] \, {\cal G} \, , \nonumber \\ 
\label{en} 
\end{eqnarray} 
%%%% End %%%%%%%%%%%%%%%%%%%%%%%%%%
where $N (\xi, {\bf k}) \equiv 1 / (e^{i \xi k \, g 
(\hat{k})} - 1)$. Since we are interested in collinear 
configuration, $\hat{{\bf k}}_n \simeq \hat{{\bf k}}_j \simeq 
\widehat{{\bf k}_n - {\bf k}_j}$ and $|{\bf k}_n - {\bf k}_j| \simeq 
k_n - k_j$. Then we have 
%%%%% Equation %%%%%%%%%%%%%%%%%%%%%%%%%%%%
\[ 
\mbox{Eq. (\ref{en})} \simeq \int d E \, \rho (E) 
\int_{- \infty}^\infty \frac{d \xi}{2 \pi} e^{i \xi E} e^{i \xi k_n 
\, g (\hat{{\bf k}}_n)} N (\xi, {\bf k}_j ) N (\xi, {\bf k}_n 
- {\bf k_j}) \, {\cal G} \, , 
\] 
%%%%% End %%%%%%%%%%%%%%%%%%%%%%%%%
which is the left-hand side of (\ref{cond2}). For other regions of 
$(p_0, k_0)$ than the one studied above, one can similarly prove 
(\ref{cond2}). 

It should be noted that, since $\rho (E)$ and $g (\hat{{\bf k}})$ 
are arbitrary functions, the resultant form for $n (k, 
\hat{{\bf k}})$ \lq\lq covers'' a wide class of functions. 
%%%%%% Acknowledgments %%%%%%%%%%%%%%%%%%%%%%
\section*{Acknowledgments}
This work was supported in part by the Grant-in-Aide for Scientific 
Research ((A)(1) (No.~08304024)) of the Ministry of Education, 
Science and Culture of Japan. 
%%%%%%%%%%%%%%%%%%%%%%%%%%%%%%%%%%%%%%%%%%%%%%%%%
%%%%%%%%%%%%%% Appendix A %%%%%%%%%%%%%%%%%%%%%%%
%%%%%%%%%%%%%%%%%%%%%%%%%%%%%%%%%%%%%%%%%%%%%%%%%
\setcounter{equation}{0}
\setcounter{section}{1}
\section*{Appendix A Absence of mass singularity \\ 
in $\Pi_\Sigma^{(n)} (Q)$} 
\def\theequation{\mbox{\Alph{section}.\arabic{equation}}}
In this Appendix, we show that $\Pi_\Sigma^{(n)} (Q)$ is free from 
mass singularities, reconfirming the result in \cite{lb}. 
Manipulation goes as follows. Substitute $F^{(n)}$, Eq. (\ref{f}), 
into (\ref{s1}) with (\ref{self}). Use the form of $g_{\mu 
\nu}^{(\mbox{\scriptsize{gauge}})}$, Eq. (\ref{cov}) or Eq. 
(\ref{t}), and forms for $\Delta_{l j}$ and $S_{l j}$ (cf. Eq. 
(\ref{hosi})). The resultant expressions may be rearranged as 
%%%%%% Equation %%%%%%%%%%%%%%%
\begin{eqnarray} 
\Pi_\Sigma^{(n)} & = & \frac{8}{3 \pi^2} \alpha \alpha_s \int 
\frac{d^{\, 4} P}{2 \pi} \int \frac{d^{\, 4} K}{2 \pi} \mbox{Tr} 
\left[ {\cal G}^{\mu \nu} 
({P\kern-0.1em\raise0.3ex\llap{/}\kern0.15em\relax} - 
{Q\kern-0.1em\raise0.3ex\llap{/}\kern0.15em\relax} ) \right] 
\nonumber \\  
& & \times \hat{g}_{\mu \nu}^{(\mbox{\scriptsize{gauge}})} \tilde{f} 
(P) (1 - \tilde{f} (P - Q)) \delta_\epsilon ((P - Q)^2) \nonumber \\ 
& & \times \left[ \left( \frac{1}{2} + f (K) \right) 
\delta_{\epsilon} (K^2) \left( \Delta_R^2 (P) \Delta_R (P - K) 
+ \Delta_A^2 (P) \Delta_A (P - K) \right) \right. \nonumber \\ 
& & \left. + \left( \frac{1}{2} - \tilde{f} (P - K) \right) 
\delta_{\epsilon} ((P -K)^2) \left( \Delta_R^2 (P) \Delta_R (K) + 
\Delta_A^2 (P) \Delta_A (K) \right) \right] \, , \nonumber \\ 
\label{A1} 
\end{eqnarray} 
%%%%% End %%%%%%%%%%%%%%%%%%%%%%%
where ${\cal G}^{\mu \nu}$ is as in (\ref{pre}) and 
$\delta_{\epsilon} (k^2) \equiv \epsilon (k_0) \delta (K^2)$ etc. 
For the Coulomb gauge, 
$\hat{g}_{\mu \nu}^{(\mbox{\scriptsize{gauge}})} = 
g_{\mu \nu}^{(\mbox{\scriptsize{t}})}$. For the covariant gauge, 
$\hat{g}_{\mu \nu}^{(\mbox{\scriptsize{gauge}})} = g_{\mu \nu} + 
\eta K_\mu K_\nu (\partial / \partial K^2)$, where $\partial / 
\partial K^2$ applies to $\delta (K^2)$, $\Delta_R (K)$ and 
$\Delta_A (K)$. Equation (\ref{A1}) is manifestly free from mass 
singularities. Mass singularity arises from the terms 
$\Delta_{R (A)}^2 (P) \Delta_{A (R)} (P - K)$ and $\Delta_{R (A)}^2 
(P) \Delta_{A (R)} (K)$. In obtaining (\ref{A1}), cancellations 
occur between those terms. 

It is also obvious from (\ref{A1}) that $\Pi_\Sigma^{(n)}$ is free 
from divergence due to infrared singularities, provided that, as $k 
\to 0^+$, $f (k, \hat{{\bf k}}) \propto k^{- n}$ with $n < 2$ and 
$\tilde{f} (k, \hat{{\bf k}}) \propto k^{- n'}$ with $n' < 2$. In 
actual computation of $\Pi$, for a propagator with soft momentum, 
one should use hard-thermal-loop resummed effective one. 
%%%%%%%%%%%%%%%%%%%%%%%%%%%%%%%%%%%%%%%
%%% Appendix B %%%%%%%%%%%%%%%%%%%%%%%%
%%%%%%%%%%%%%%%%%%%%%%%%%%%%%%%%%%%%%%%
\setcounter{equation}{0}
\setcounter{section}{2}
\section*{Appendix B Computation of the singular part of 
$\Pi_\Sigma^{(p)} (Q)$} 
\def\theequation{\mbox{\Alph{section}.\arabic{equation}}}
Here, we compute the singular part of the \lq\lq pinch'' 
contribution $\Pi_\Sigma^{(\mbox{\scriptsize{p}})} (q_0, {\bf q} = 
{\bf 0})$. Substituting (\ref{p}) into (\ref{s1}) with (\ref{self}) 
and using (\ref{pre}) and (\ref{tr}), and the forms for 
$\Delta_{l j}$ and $S_{l j}$, we have 
%%%%%%  Equation %%%%%%%%%%%%%%%
\begin{eqnarray*} 
\Pi_\Sigma^{(p), \, \mbox{\scriptsize{sing}}} & = & - 
\frac{128}{3 \pi} \alpha \alpha_s Q^2 \int \frac{d^{\, 4} P}{2 \pi} 
\int \frac{d^{\, 4} K}{2 \pi} G^{(\mbox{\scriptsize{gauge}})} 
(1 - \tilde{f} (P - Q)) \delta_\epsilon (K^2) 
\delta_\epsilon ((P -K)^2) \\ 
& & \times \delta_\epsilon ((P - Q)^2) \, {\bf P} \frac{1}{P^2} [ f 
(K) \tilde{f} (P - K) 
- \tilde{f} (P) ( 1 + f (K) - \tilde{f} (P - K) ) ] \, , 
\end{eqnarray*} 
%%%%%% End %%%%%%%%%%%%%%%%%%%%%%
where 
%%%%% Equation %%%%%%%%%%%%%%%%
\begin{eqnarray*} 
G^{(\mbox{\scriptsize{Fey}})} & = & \frac{k_0}{q_0} \, , \\ 
G^{(\mbox{\scriptsize{Cou}})} & = & \frac{(q_0 - k_0)^2 + 
k_0^2}{2 q_0 k_0} \, . 
\end{eqnarray*} 
%%%%% End %%%%%%%%%%%%%%%%%%%%%%%
Here use has been made of $p_0 = q_0 / 2$, which comes from 
$(P - Q)^2 = 0$. Making the change of variable $P \to P + K$, we 
extract the mass-singular part, 
%%%%%%  Equation %%%%%%%%%%%%%%%
\begin{eqnarray} 
& & \delta (P^2) \int d^{\, 4} K \delta (K^2) \, {\bf P} 
\frac{1}{(P + K)^2} \nonumber \\ 
& & \hspace*{4ex} = \pi \delta (P^2) \int d k \, k^2 \int 
d k_0 \, \delta (K^2) \int^{1 - \epsilon_y}_{- 1 + 
\epsilon_y} d (\hat{{\bf p}} \cdot \hat{{\bf k}}) \frac{1}{p_0 k_0 - 
{\bf p} \cdot {\bf k}} \nonumber \\ 
& & \hspace*{4ex} \to \frac{\pi}{p} \delta (P^2) \int d k \, 
k \int d k_0 \, \epsilon (p_0 k_0) \delta (K^2) \ln 
\frac{1}{\epsilon_y} \, . 
\label{A3} 
\end{eqnarray} 
%%%%% End %%%%%%%%%%%%%%%%%%%%%%%
Using (\ref{bunpu}), cutting off the infrared region, $k \; (\equiv 
\kappa z) \geq \epsilon_z \kappa$, and changing the integration 
variable suitably, we arrive at the final form. As it should be, 
$\delta \Pi_\Sigma^{(p), \, \mbox{\scriptsize{sing}}} = 
\Pi_\Sigma^{(p), \, \mbox{\scriptsize{sing}}} \mbox{\hspace*{.4mm}} 
\rule[-2.5mm]{.14mm}{6.2mm} \raisebox{-2.45mm}{\scriptsize{$\; 
\mbox{\scriptsize{Fey}}$}} - \Pi_\Sigma^{(p), \, 
\mbox{\scriptsize{sing}}} \mbox{\hspace*{.4mm}} 
\rule[-2.5mm]{.14mm}{6.2mm} \raisebox{-2.45mm}{\scriptsize{$\; 
\mbox{\scriptsize{Cou}}$}}$ is equal to 
$- \Pi_V^{\mbox{\scriptsize{sing}}}$, which has been evaluated in 
\cite{lb}. There is a missing term though in $\Pi_V^{\, 
\mbox{\scriptsize{sing}}}$ in \cite{lb}, 
%%%%% Equation %%%%%%%%%%%
\[ 
- \frac{32}{3 \pi} \alpha \alpha_s \kappa^2 (\tilde{n} (\kappa))^2 
\ln \frac{1}{\epsilon_y} \int_{\epsilon_y}^1 d z \, \frac{1 - z}{z} 
\, , 
\] 
%%%%%% End %%%%%%%%%%%%%%%%
where $\kappa = q_0 / 2$. The form for $\Pi^{\, 
\mbox{\scriptsize{sing}}}$, being gauge independent, reads 
(\ref{final}) in the text.   
%%%%%%%%%%%%%%%%%%%%
%%% REF %%%%%%%%%%%%%%%%%%%%%%
%%%%%%%%%%%%%%%%%%%%%%%%%%%%%%%%%%%

\end{document}